\documentclass[a4paper,fleqn,usenatbib,letter]{mnras}

\usepackage{newtxtext,newtxmath}

\usepackage[T1]{fontenc}
\usepackage{ae,aecompl}

\usepackage{graphicx}	
\usepackage{amsmath}	
\usepackage{amssymb}	

\title[Does the Void-in-Cloud Process Matter?]{Void Formation: Does the Void-in-Cloud Process Matter?}

\author[H. Y. J. Chan et al.]{
Hei Yin Jowett Chan,$^{1}$\thanks{E-mail: jchan@astr.tohoku.ac.jp}
Masashi Chiba,$^{1}$
Tomoaki Ishiyama$^{2}$
\\
$^{1}$Astronomical Institute, Tohoku University, Aoba-ku, Sendai, 980-8578, Japan\\
$^{2}$Institute of Management and Information Technologies, Chiba University, 1-33, Yayoi-cho, Inage-ku, Chiba, 263-8522, Japan\\
}

\date{Accepted XXX. Received YYY; in original form ZZZ}

\pubyear{2019}

\begin{document}
\label{firstpage}
\pagerange{\pageref{firstpage}--\pageref{lastpage}}
\maketitle

\begin{abstract}
We investigate the basic properties of voids from high resolution, cosmological N-body simulations of $\Lambda$--dominated cold dark matter ($\Lambda$CDM) models, in order to compare with the analytical model of Sheth and van de Weygaert (SvdW) for void statistics. For the subsample of five dark matter simulations in the $\Lambda$CDM cosmology with box sizes ranging from 1000$\,h^{-1}$Mpc to 8$\,h^{-1}$Mpc, we find that the standard void--in--cloud effect is too simplified to explain several properties of identified small voids in simulations. (i) The number density of voids is found to be larger than the prediction of the analytical model up to $2$ orders of magnitude below 1$\,h^{-1}$Mpc scales. The Press-Schechter model with the linear critical threshold of void $\delta_\text{v} = -2.71$, or a naive power law, is found to provide an excellent agreement with the void size function, suggesting that the void-in-cloud effect does not suppress as much voids as predicted by the SvdW model. (ii) We then measured the density and velocity profiles of small voids, and find that they are mostly partially collapsing underdensities, instead of being completely crushed in the standard void--in--cloud scenario. (iii) Finally, we measure the void distributions in four different tidal environments, and find that the void--in-void effect alone can explain the correlation between distribution and environments, whereas the void--in--cloud effect is only weakly influencing the abundance of voids, even in filaments and clusters.
\end{abstract}



\section{Introduction}

Although the $\Lambda$CDM model has successfully explained observations on large scales, it is currently facing a number of challenges on galactic and sub--galactic scales. Examples are the Missing Satellite Problem and the Void Phenonmenon, where the former describes the lack of observed dwarf galaxies within an overdense structure such as the Milky Way halo, and the latter describes the lack of galaxies within an underdense structure, the Local Void \citep{Peebles2010,Hoeft2006,Tavasoli2013}. However, we must stress that most of the studies only considered voids with sizes larger than 1 $h^{-1}$Mpc, whereas the discrepancy with $\Lambda$CDM occurs on smaller scales \citep{Bullock2017}.

One of the solutions to the $\Lambda$CDM problem is to allow dark matter particles having non--negligible thermal velocity during the epoch of structure formation. This alternative model, known as the Warm Dark Matter (WDM), is proposed to solve the problem regarding the overabundance of subhalos, since its free streaming effect can suppress halo formation on small scales. For voids, \citet{Tikhonov2009} suggested that the WDM model provides an easy fix to the observed abundance of small voids within the Local Volume, but unlike halo formation under the WDM Universe which is studied in--depth by numerous researchers, void statistics in the WDM Universe is rarely discussed, with a few exceptions \citep{Reed2015,Yang2015}. It is possible, nevertheless, to use an analytical void model developed by \citet{Sheth2004} to reveal the small--scale behaviour of void distribution in the WDM Universe. 
   
\citet{Sheth2004} (SvdW) formulated an analytical void size function through the excursion set approach and suggested that the formation of small voids do not only depend on a sufficiently underdense region in the initial condition, but also the so--called void--in--cloud effect: a void embedded in a sufficiently overdense region will be crushed out of existence. In this case, the abundance of small voids is greatly reduced in the CDM Universe. Since it is usual to study the halo mass function in the WDM Universe by applying a simple truncated linear power spectrum to the Extended--Press Schechter (PS) model \citep{Schneider2013}, we could use the same approach as the SvdW model and found that the void size function behaves distinctively different from the halo function, because of the void--in--cloud effect. 

Although the SvdW model becomes the foundation of many other modified models for void abundances, such as the volume conserving model \citep{Jenning2013}, voids in modified gravity model \citep{Clampitt2013}, Eulerian void--in--cloud formulation \citep{Paranjape2012}, and many others \citep{Achitouv2013,Lam2014}, it is not clear if we can safely adopt the void--in--cloud process in the calculation when we lack a full understanding of it. 

The motivation of the paper is to investigate the possibility of using a void size function to distinguish between the CDM and WDM cosmology. Along the way, a distinctive behaviour is found on scales below 1 $\,h^{-1}$Mpc, where the small--scale issues of $\Lambda$CDM become relevant, but the behaviour is caused by the rarely discussed void--in--cloud process. Therefore, we here study the void distribution on scales below 1 $\,h^{-1}$Mpc, where both the void--in--cloud process and $\Lambda$CDM problems become relevant, using both the theoretical approach and cosmological simulations in the $\Lambda$CDM Universe. The measured abundance of small voids should reveal if the void--in--cloud effect plays an important role in the void formation process or not. 

In this paper, we identify spherical and aspherical voids in Multi-Dark simulation\footnote{\url{https://www.cosmosim.org}} \citep{Klypin2016}, Phi-0 \citep{Ishiyama2016} and Phi-1 simulations\footnote{\url{http://hpc.imit.chiba-u.jp/~ishiymtm/db.html}} \citep{Ishiyama2019} based on Voronoi tessellation method. We find that the assumed void--in--cloud effect in the SvdW model is too simplified to explain several properties of small voids with sizes below $1\,h^{-1}$Mpc. Note that the void--in--cloud effect provides a more direct impact on dark matter voids, so we are not considering galaxy voids in this work, although it is still possible with the consideration of galaxy bias \citep{Furlanetto2005}.

This paper is organized as follows. In Section 2, we briefly review the analytical void distribution predicted by the SvdW model. In Section 3, we present the void distribution in the WDM Universe in comparison with halo distribution. In Section 4, we summarize the simulation and void finding method that we used in this work. The major results on the property of small voids are presented in Section 5. Finally, we discuss our result in Section 6 and conclude  in Section 7. 
\begin{table*}
	\caption{Details of the simulations\label{tab:1}. Here $N$, $N_{\text{sub}}$, $m_\text{p}$ and $\epsilon$ are the number of particles, the subsample, mass resolution and gravitational softening length respectively. (1) Ishiyama et al. 2016 (2) Ishiyama $\&$ Ando 2019 (3) Klypin et al. 2016}
	\label{tab: 1}
	\begin{tabular}{lcccccc}
		\hline
		\hline
		Name & box size & $N$ & $N_{\text{sub}}$ & $m_\text{p}$ & $\epsilon$ &Ref\\
		&($h^{-1}$Mpc)&	  &                  &$(h^{-1}\text{M}_{\odot})$& ($h^{-1}$kpc) & \\
		\hline
		Phi-0 & $8$ 	 & $2048^3$ & $[1.07, 8.59] \times 10^7$ & $5.13\times 10^3$ & 0.12 & $1$ \\
		Phi-1 & $32$ 	 & $2048^3$ & $8.59 \times 10^7$ & $3.28\times 10^5$ 	& 0.48 & $2$ \\
		VSMDPL& $160$  & $3840^3$ 	& $5.66 \times 10^7$ & $6.20\times 10^6$ & 1 & $3$\\
		SMDPL & $400$  & $3840^3$ 	& $5.66 \times 10^7$ & $9.63\times 10^7$ & 1.5 & $3$\\
		MDPL2 & $1000$ & $3840^3$ 	& $5.66 \times 10^7$ & $1.51\times 10^9$ & 5 & $3$\\
		\hline
	\end{tabular}
\end{table*}
\section{Excursion Set Formalism of Voids}
The PS formalism  is used for the distribution of dark matter halos, which has been proven to be a success when comparing to cosmological simulations \citep{Bond1991,Press1974,Warren2006,Zenter2007}. SvdW proposed an analogous approach to model the distribution of voids by considering the spherical expansion model and excursion set formalism with the consideration of the two barrier problem. Here we present a brief review on the SvdW model.

Suppose that the linear perturbation field $\delta(\vec{x})=(\rho(\vec{x})-\rho_\text{m})/\rho_\text{m}$ can be smoothed and represented by its Fourier transform $\delta(\vec{k})$.
To characterize the perturbation field in the statistical perspective, it is useful to introduce the variance of the field
\begin{equation}
\sigma^2(R) = \dfrac{1}{2\pi^2} \int P(k) W^2(k,R) k^2dk ,
\end{equation}
where $P(k)$ is the matter power spectrum, $W(k,R)$ is the Fourier transformed spherical top-hat filter function, and the smoothing scale $R$ corresponds to the radius of a spherical region. 

According to the spherical evolution model, underdense or overdense objects evolve non-linearly after shell crossing, which occurs at the linear critical density $\delta_\text{v} = -2.71$ for voids and $\delta_\text{c} = 1.686$ for collapsed objects in an Einstein de-Sitter universe. Since changes of these density thresholds in the $\Lambda$CDM universe are small, we use these values throughout this paper. With these two thresholds, we can formulate the distribution of voids in linear theory as
\begin{equation}
\dfrac{dn}{d\ln R_\text{L}}=\dfrac{S f(S,\delta_\text{c},\delta_\text{v})}{V(R_\text{L})}\dfrac{d\ln S}{d\ln R_\text{L}} ,
\end{equation}
where $S \equiv \sigma^2$ and $V(R) \equiv 4\pi R^3/3 $. The excursion set formalism of collapsed objects defines $f(S)$ as the fraction of uncorrelated random walks that first cross the barrier $\delta_\text{c}$ on the $\delta(\sigma^2)$ map, 
\begin{equation}
f_{\text{PS}}(S,\delta_\text{c})=\dfrac{\delta_\text{c}}{\sqrt{2\pi S^3}}\exp[-\delta^2_\text{c}/(2S)].
\end{equation}
The excursion set formalism of the void requires two barriers, $\delta_\text{c}$ and $\delta_\text{v}$. This is a result of taking into account of the void--in--void and void-in-cloud processes. The latter process occurs when voids embedded within a larger overdense region at earlier epoch are squeezed out of existence at later epoch, resulting in the decrease of small voids. The barrier crossing distribution of void becomes
\begin{equation}
f(S,\delta_\text{c},\delta_\text{v}) = \sum_{j=1}^{\infty}\dfrac{j^2\pi D^2}{(\delta^2_\text{v}/S)}\dfrac{\sin(j\pi D)}{j\pi}\exp\left[-\dfrac{j^2\pi^2 D^2}{2(\delta^2_\text{v}/S)}\right],
\end{equation}
where $D = |\delta_\text{v}|/(\delta_\text{c}+|\delta_\text{v}|)$. $f_\text{PS}(S,\delta_\text{c})$ and $f(S,\delta_\text{c},\delta_\text{c})$ can also be understood as the probability distribution of halo and void formations, respectively. In addition, the spherical expansion model predicts that the non-linear radius of a void is related to the linear radius by an expansion factor, $R \approx 1.7 R_\text{L}$. Then, the void size function becomes 
\begin{equation}
\dfrac{dn}{d \ln R}=\left(\dfrac{dn}{d \ln R_\text{L}}\right)\Big|_{R_\text{L} = R/1.7}.
\end{equation}

Note that SvdW is not an accurate model of the void size function, because it uses approximations such as the spherical symmetry where realistic voids are mostly aspherical, and uncorrelated random walk due to the use of a sharp--k filter. \citet{Paranjape2012} proposed using an Eulerian approach to account for voids that are only partially squeezed but not completely crushed by the void--in--cloud effect. We will see later how it is relevant to our result.
\section{Void Function in the WDM Universe} 
Although the SvdW model suffers from a few approximations, it can still provide us some insight into the behaviour of void distribution in both the CDM and WDM Universe. The size function depends on the linear power spectrum of the density perturbations, which has the following proportionality at current epoch $P(k) \propto k^nT^2(k)$ where  $n$ is the spectral index. An analytical expression for transfer function $T(k)$, including the free streaming effect from WDM particles, can be derived from the coupled linearized Einstein-Boltzmann equations, but also can be expressed by fitting functions obtained from Boltzmann code simulation. For the CDM and WDM Universe, we use that provided by \citet{Bardeen1986}. For the WDM Universe, \citet{Bode2001} proposed the following

\begin{subequations}
	\begin{align}
	&T_\text{WDM}(k)=[1+(\alpha k)^{2\nu}]^{-5/\nu}\\
	&\alpha=0.048\left(\dfrac{\Omega_m}{0.4}\right)^{0.15} \left(\dfrac{h}{0.65}\right)^{1.3}
	\left(\dfrac{m}{\text{keV}}\right)^{1.15} \left(\dfrac{1.5}{g}\right)^{0.29},
	\end{align}	
\end{subequations}
where $g$ is the number of degree of freedom and we set it as $g=1.5$ for simplicity. $m$ is the mass of a WDM particle, which is currently unknown but one of the important tasks is to constrain this parameter. The current lower limit is $\sim$2keV \citep{Hayashi2015, Viel2005}. 

\begin{figure*}
		\includegraphics[scale=0.8]{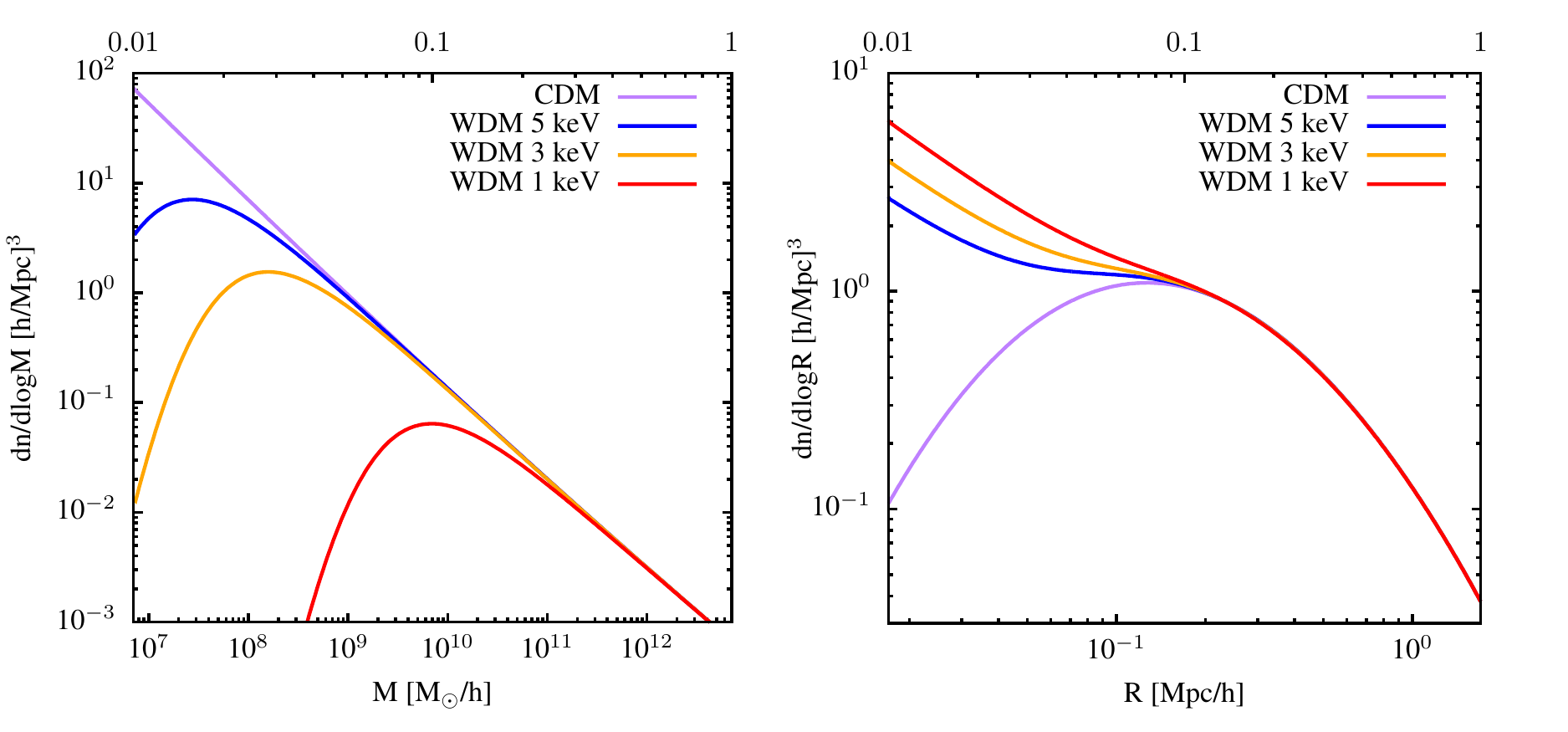}
		\centering
		\caption{
	Halo mass function on the left predicted by PS model and void size function SvdW
	model on the right. Top and bottom x--axis denote the linear radii and non--linear property of the
	structure.
			}
		\label{fig: psvdw}
\end{figure*}

Figure \ref{fig: psvdw} demonstrates the behaviour of halo mass and void size functions in both the CDM and WDM Universe within the same range of linear radius. The mass of the WDM particle ranges from 1keV to 5keV. 
The standard PS model is known to fail dramatically in WDM cosmology on scales below the half--mode scale due to the use of a spherical top--hat filter, so we instead replace it with a sharp--k filter and introduce an appropriate mass assignment $M = \rho_\text{m}4\pi (2.7R_\text{L})^{3}/3$ suggested by \citet{Schneider2013}. As the left panel of Figure \ref{fig: psvdw} shows, the halo abundance keeps increasing to smaller scales for the CDM cosmology, but it is suppressed in the WDM cosmology as expected. In the case of voids, we observe an opposite behaviour. The abundance of small voids is significantly suppressed in the CDM cosmology, while the abundance increases in the WDM cosmology. 

The prediction can be explained by the void--in--cloud effect. Small voids are mostly crushed out of existence by an overdense shell in the CDM Universe, but overdense structures become fewer in the WDM Universe, which reduce the occurrence of the void--in--cloud process and allow more formation of small voids. The prediction also implies that the void--in--cloud effect seems to be primarily caused by overdense structures below the half--mode scale. However, the use of a spherical top--hat filter causes the standard PS model to fail below half--mode scale, so a similar conclusion may also be made for the SvdW model. We suggest that the question of whether the void--in--cloud process is suppressed in WDM cosmology or not could be resolved by studying voids in WDM simulations. If it is true, the unique behaviour of void distribution may allow us to develop an independent new probe of the property of WDM. 

Although the void--in--cloud process is suggested as the cause of the behaviour of the void function in Figure \ref{fig: psvdw}, the physics of the process is still rather uncertain. Therefore, before we advance to study such as the abundance of small voids in WDM simulation or observational statistics of small voids \citep{Furlanetto2005}, we must first clarify the happening of the void--in--cloud process, which is the major purpose of this paper.

\section{Simulations and Void Finder}
To compare with the SvdW model, we adopt data from five cosmological N-body simulations with box sizes ranging from 1000 $h^{-1}$Mpc down to 8 $h^{-1}$ Mpc, where two of them are Phi-0 and Phi-1 simulations \citep{Ishiyama2016,Ishiyama2019} and the other three are the MultiDark-Planck simulations \citep{Klypin2016}. The wide ranges of box sizes allow identification of voids with sizes spanning over $3$ orders of magnitudes. All simulations are performed under Planck cosmology where the cosmological parameters are $\Omega_\text{m} \approx 0.31$, $\Omega_\Lambda\approx0.69$, $\Omega_\text{b}=0.048$, $n_\text{s}=0.96$, $\sigma_8\approx0.83$ and $h\approx0.68$. Snapshots of the simulations at current epoch, which originally contain $N$ number of dark matter particles of mass $m_\text{p}$, are randomly subsampled into $N_{\text{sub}}$ particles. The choice of $N_{\text{sub}}$ is to allow maximum resolution and minimum time consumption during the void finding processes. The details of the simulations can be found in Table~\ref{tab: 1}.

The SvdW model assumes voids to be top-hat spherical underdense regions with non-linear average density $\rho_\text{v}=0.2\bar{\rho}$, where $\bar{\rho}$ refers to the mean particle density of the subsampled simulation. In order to make a consistent comparison of void size function, we must also assume voids to be spherical during the void finding process. However, since taking account of apshericity could also provide us a more realistic picture of voids, our void finding method will identify both spherical and aspherical voids in the simulations. In general, we expect apsherical voids to be larger, but it will also be interesting to see how the void distribution changes regarding different assumption of shape.

Our void finder uses the untrimmed outputs from the public void finding algorithm VIDE\footnote{\url{http://www.cosmicvoids.net}} \citep{Sutter2015}, which is heavily based on ZOBOV \citep{Neyrinck2008}. Since VIDE can output all aspherical parent Voronoi voids, no further processing is required to generate a catalog of aspherical void. Here we only consider parent voids, in order to take account of the void--in--void effect. However, the size of aspherical voids strongly depends on the zone merging process controlled by a merging criterion. The default setting of VIDE assumes the criterion to be $\rho_\text{link}<0.2\bar{\rho}$, but this choice of criterion lacks a theoretical support. Furthemore, the default setting fails to measure a continuous void size function with different boxsizes. For example, the default VIDE finds an extremely large parent void in Phi--0 and Phi--1 occupying almost the entire simulation, but, at the same time, plentiful small voids still survive through the zone merging process. The resulting size function becomes truncated only on the intermediate scales, but not on smaller scales. To resolve the issue, the usual practice is to treat the threshold as a free parameter \citep{Nadathur2015}. After some experimentation, we found that the following merging criteria $\rho_\text{link} < [0.032, 0.04, 0.1, 0.15, 0.24] \bar{\rho}$ for Phi--0, Phi--1, VSMDPL, SMDPL and MDPL2 simulations, respectively, allows us to measure a continuous power--like void size distribution. Remind that if the threshold is too small, no zone merging will occur, resulting in voids without the void--in--void consideration. We claim that it will be an essential work to develop a void finding method that is in consistent regardless of boxsizes of simulation, but we will leave this to future study.
  
For spherical voids, we follow \citet{Jenning2013} to construct our own spherical void catalog by growing spheres at the volume-weighted center of each Voronoi voids. Then, we iteratively include particles one by one until the maximum radius at which the average density reaches the non-linear critical density $0.2\bar{\rho}$. Lastly, to avoid double counting of voids, we pick up the void with the most underdense core particle if overlap occurs. Note that our void finder does not completely follow the criteria in \citet{Jenning2013}. We do not consider a minimum and maximum cutoff radius before growing the spheres but after it in order to remove spurious voids that are clearly due to the limitation of resolution. For both aspherical and spherical voids, we only consider voids with core particles' densities that are greater than $0.2\bar{\rho}$ to reject statistically insignificant voids \citep{Neyrinck2008}. 

\begin{figure*}
	\includegraphics[scale=0.8]{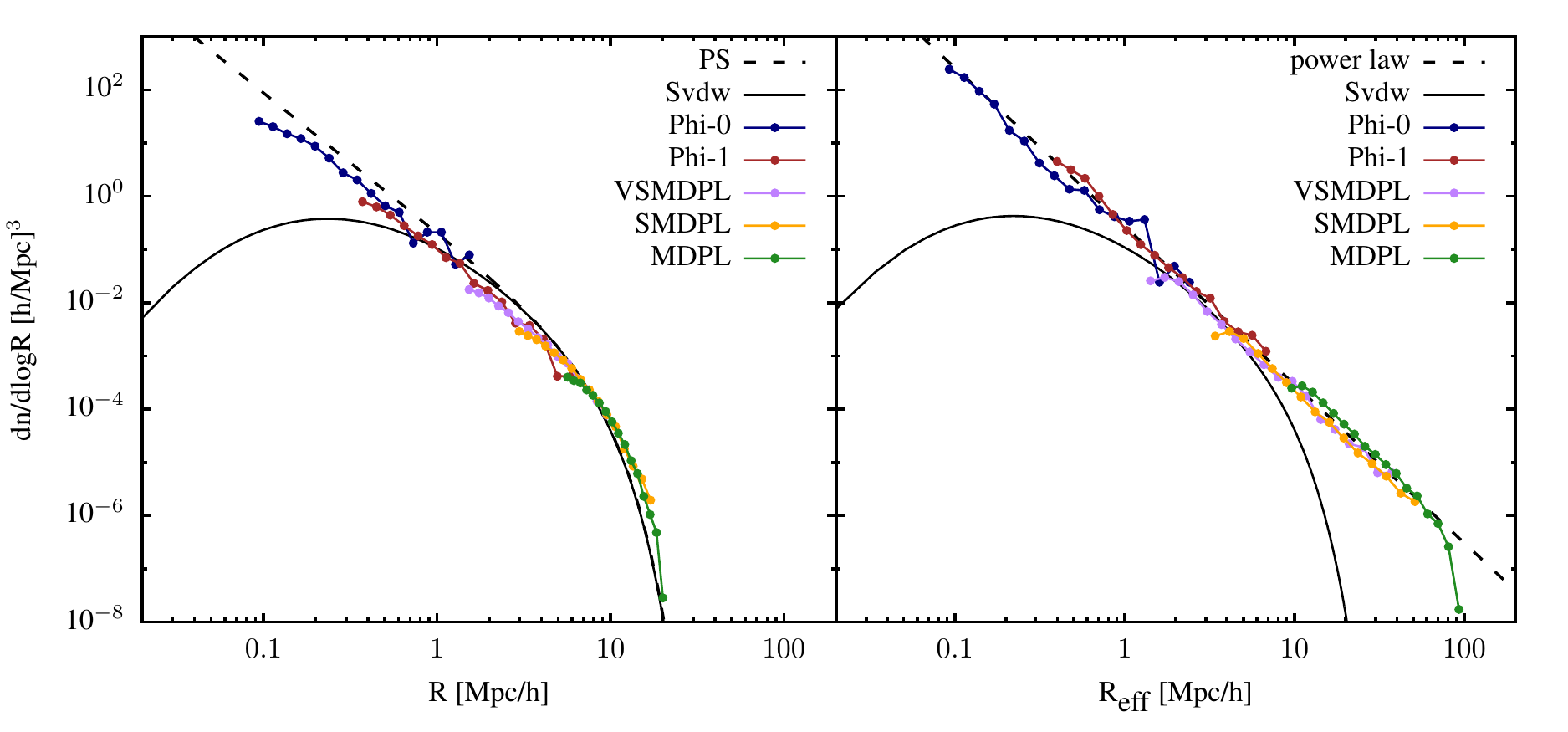}
	\centering
	\caption{
	Spherical (left) and apsherical (right) void size distributions in subsampled simulations vs. theoretical models. Different linespoints correspond to simulations with box size 8 $h^{-1}$Mpc (blue), 32 $h^{-1}$Mpc (brown), 160 $h^{-1}$Mpc (purple), 400 $h^{-1}$Mpc (orange) and 1000 $h^{-1}$Mpc (green). Black solid line refers to the two barrier excursion set model. Black dash line refers to the one barrier
	excursion set model using the linear critical density of void $\delta_\text{v} = -2.71$ for spherical void and a navie power law for aspherical void.
	}
	\label{fig: savsf}
\end{figure*}
\begin{figure*}
	\includegraphics[scale=0.8]{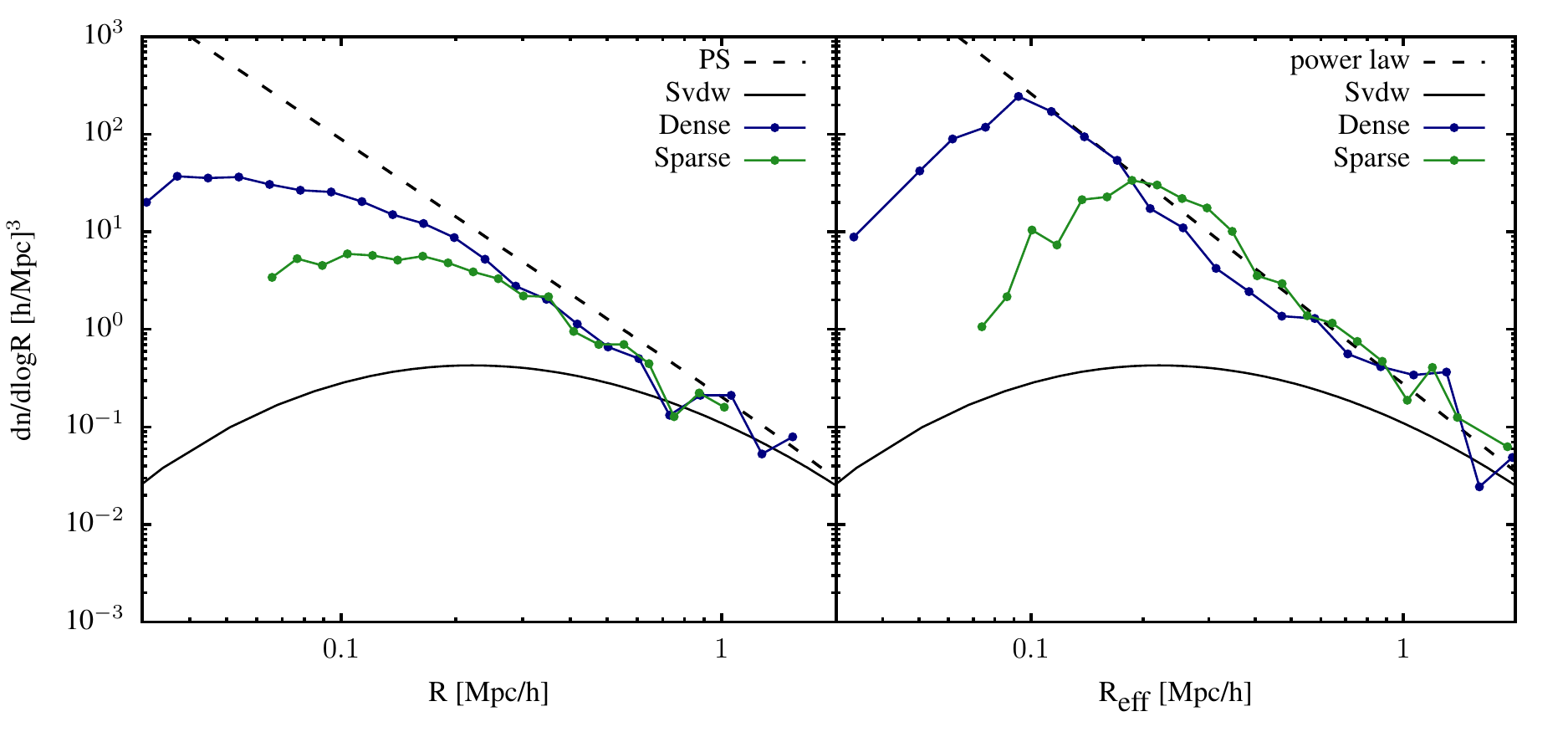}
	\caption{
		Spherical (left) and spherical (right) void size distributions in subsampled simulations vs theoretical models. Blue and green linespoints correspond to the measurement in subsample of Phi-0 simulation with $N_{\text{sub}}= 1.07 \times 10^7$ and $8.58 \times 10^7$ respectively. Black solid line refers to the two barrier excursion set model. The black dash line refers to the one barrier excursion set model using the linear critical density of void $\delta_\text{v}=-2.71$ for spherical void, and a naive power law for aspherical void. 
	}
	\label{fig: savsfss}
\end{figure*}
\section{Property of Small Voids}
\subsection{Void Size Function}
Figure \ref{fig: savsf} shows the spherical (left) and aspherical (right) void size functions measured from all five simulations comparing with the SvdW model (black solid). The abundance of spherical voids in simulation is found to be consistent with the SvdW model on large scales down to $R \sim 1\,h^{-1}$Mpc, but the smaller voids, where the void-in-cloud process is supposed to take  place, are found to be more abundant than the model with deviation up to 2 orders of magnitudes.  We attempt to modify the standard SvdW model by increasing the critical overdense threshold $\delta_\text{c}$ to reduce the void-in-cloud effect, which results in a better fit to the simulation. Then, we directly use the PS model but replacing $\delta_\text{c}$ with the critical underdense threshold $\delta_\text{v}$. As shown on the left panel of Figure \ref{fig: savsf} (black dash), it provides a much better fit than the SvdW on small scales. This result suggests that voids in simulation are rarely fully crushed by the void-in-cloud effect. 

We suspect if the overabundance on small scales, below $R \sim 1 \, h^{-1}$Mpc, is caused by the assumption of sphericity in the void finder. We measure aspherical void size function, shown on the right panel of Figure \ref{fig: savsf} where $R_\text{eff}$ is the effective radius of the aspherical void. We confirm that there is still an overabundance of small voids that keeps increasing to smaller scale. The aspherical void function can be described by a naive power law, which is expected to disagree with the SvdW model. Moreover, the distribution falls off relative to the power law at $\sim 90 \,h^{-1}$Mpc; whereas the spherical void distribution falls off at a smaller radius $\sim 7\,h^{-1}$Mpc, because spherical voids tend to fragment apsherical voids into smaller sizes, but, most importantly, the apshiercal void function does not fall off on small scales. Since both spherical and aspherical voids show a continually increasing distribution on small scales, it strongly suggests that the void--in--cloud process is a rare occurrence in simulation.

One of the drawbacks is the subsampling effect on the void size distribution. Since our concern is only on small scales, we apply our void finder in a sparse and dense subsample of Phi-0 simulation, corresponding to $N_\text{sub} = 1.07$  and 8.58 $\times$ 10$^7$ respectively. As shown in Figure \ref{fig: savsfss}, smaller voids are resolved and their abundance continues to increase, which suggests that increasing the size of sub-sample is not a remedy to the inconsistency between SvdW model and simulation.

\subsection{Density and Velocity Profiles}
\begin{figure*}
	\includegraphics[scale=1]{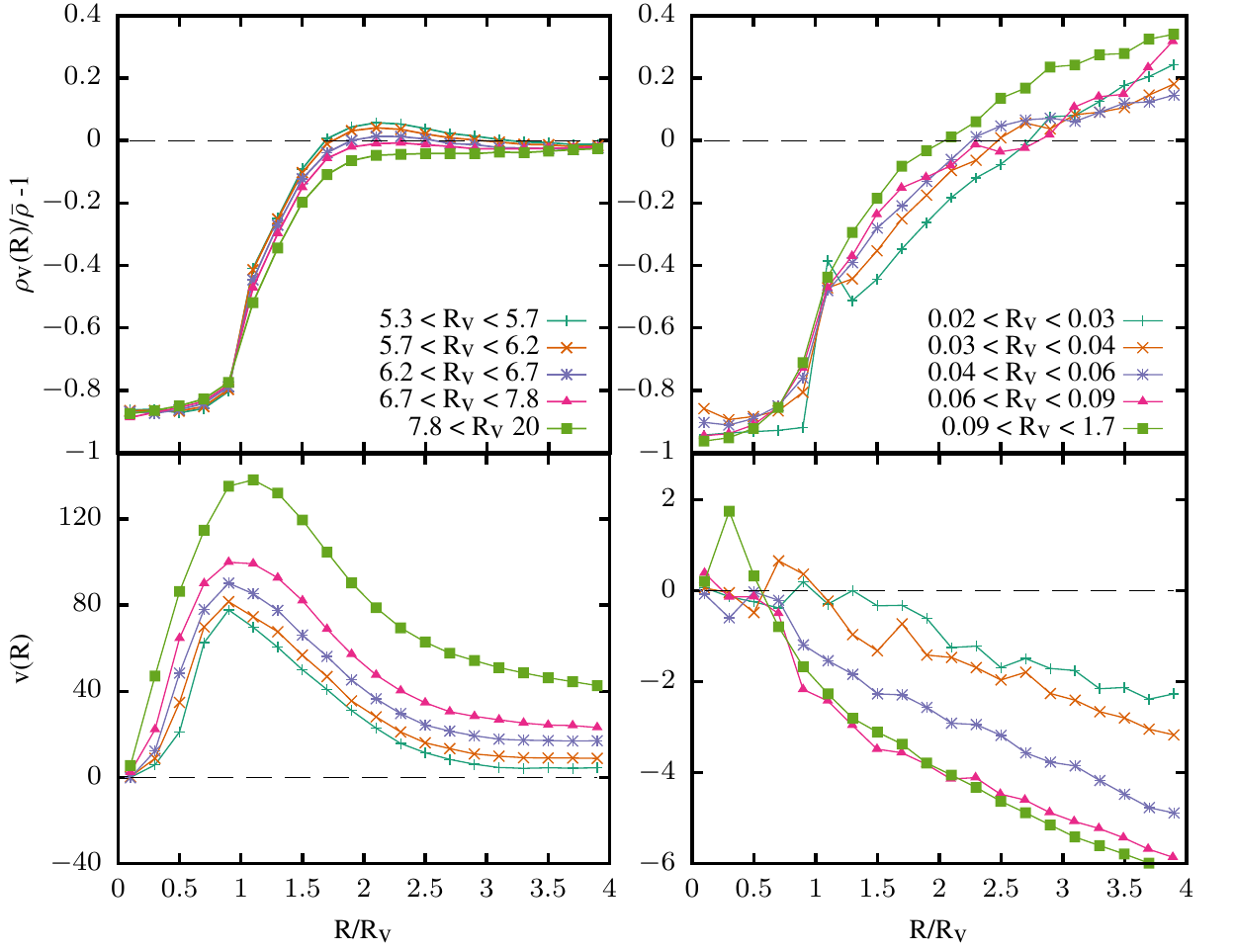}
	\centering
	\caption{
	The left and right panels show the profiles of large voids identified in the MDPL2
	simulation and small voids identified in Phi--0 simulation, respectively. The upper and lower panels correspond to the averaged density and velocity profiles in 5 contiguous bins in void radius
	respectively. $R_\text{v}$ denotes the radius of a spherical void identified by the void finding algorithm.
	}
	\label{fig: dvprof}
\end{figure*}

So far, we assume that our void finder identifies correct voids on all scales, but our question is if the small voids are mostly numerical artefacts during the void finding process. To answer it, we must look for signs of spurious voids only on small scales that are distinctively different from those larger than $1\, h^{-1}$Mpc. An intuitive approach is to make a comparison of density and velocity profiles, which are found to strongly depend on the size of voids in the sample \citep{Hamaus2014}. We follow the same approach and calculate the averaged density and velocity profiles. Figure \ref{fig: dvprof} shows the resulting profiles of voids in MDPL2 (left column) and Phi--0 (right column) binned into 5 different void radius. As expected, large voids in MDPL2 are extremely underdense at the center and the void particles are moving away from the center, proving that they are underdense and expanding structure like a standard definition of void. In general, their dependence on the size of voids also agree with the findings by \citet{Hamaus2014}, where the compensated wall becomes more pronounced and the peak velocity decreases as void decreases their size.

However, their work stopped at voids with effective radius down to 8.7 $h^{-1}$Mpc, but our sample of voids can go lower than that. The right panel of Figure \ref{fig: dvprof} shows the profiles of small voids found in Phi--0 simulation. Again, a deeply underdense regions are found at the center of these voids, proving that they are indeed voids, but unlike large voids, the density profiles of small voids increase linearly without any sign of falling back to the
global density at large radius. Compensated wall, which is expected for small voids, does not appear at these extremely small underdense structures. For velocity profile, we observe a linearly decreasing profile down to negative velocity without a peak, meaning that the outer region of small voids is a slowly collapsing structure. The void--in--cloud effect claims that voids within an overdense structure will eventually collapse completely, but the realistic case reveals that the infalling structure is not quick enough to allow full collapse at current epoch, so these voids are only partially collapsed. In fact, this is not a new idea, which is first suggested by \citet{Paranjape2012} in an attempt to integrate the partial collapsing case in the SvdW model, so we here provided an evidence for this occurrence.

\begin{figure*}
	\includegraphics[scale=0.2]{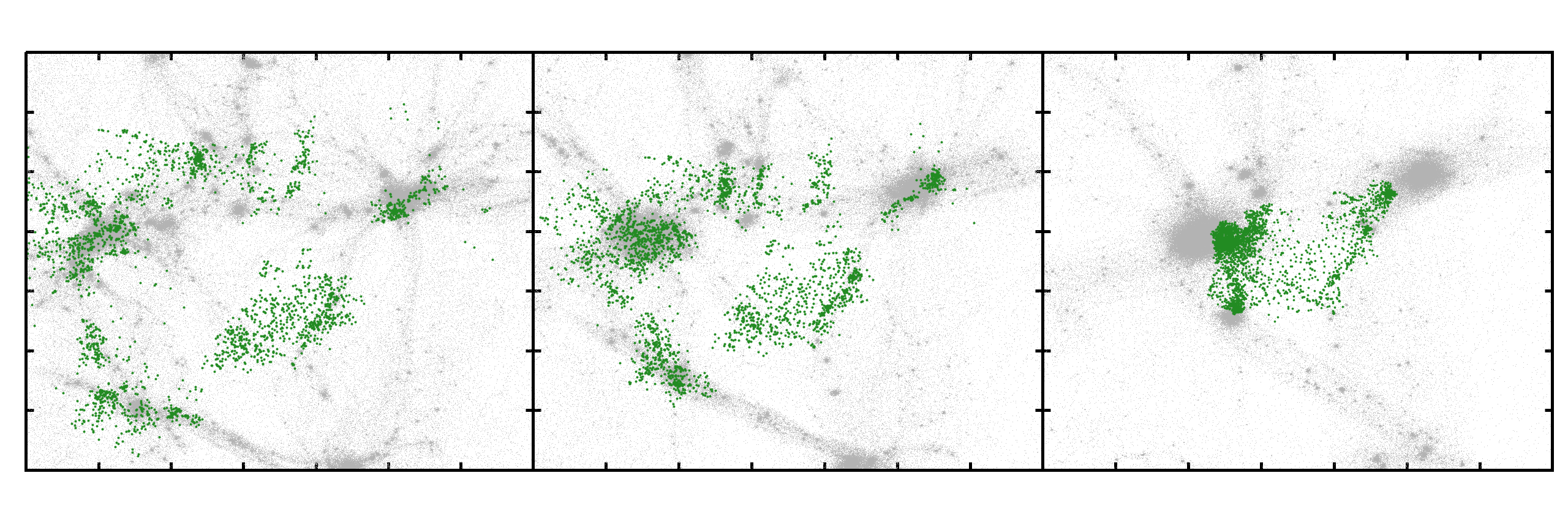}
	\centering
	\caption{
		A void with $R_\text{eff} = 0.2 \, h^{-1}$Mpc in Phi--0 simulation at redshift $z= 2,1,0$ from left to right.
	}
	\label{fig: vz}
\end{figure*}

Figure \ref{fig: vz} presents the evolution of a small aspherical void with effective radius $R_\text{eff} \sim 0.2 \,h^{-1}$Mpc from redshift $z=2$ to $z=0$ from left to right. This particular void is composed of three virialized halos at the corners with faint filaments forming a box shape identified by ZOBOV/VIDE, which is a typical example of a void--in-cloud scenario. The void particles (green) belong to the void identified at current epoch, and we trace back the corresponding particles in higher redshift snapshots. The particles at higher redshift are spreaded widely forming a gradually larger volume up to $z=2$. This void is a standard example of the void--in--cloud scenario, but only partially collapses and survives at current epoch. 

It is clear that the velocity profile provides us a distinctive difference between large and small voids, which are defined by their outer outflowing and infalling structures respectively. Taking advantage of this distinction, we could prune away spurious voids and attempt to bring the measured void size function closer to the SvdW prediction. We tried two additional pruning methods but both do not bring full consistency on all scales for spherical void distribution. First, we define and prune away spurious voids as those having infalling speed over a negative threshold within a certain normalized radius, or, in other words, those having a more dominant infalling structure. Although the corrected spherical void distribution can be forced to match with the SvdW prediction on small scales, it fails to match on the intermediate scales. Another attempt is to define spurious voids as those having more isotropic velocities, which can be seen in Figure \ref{fig: dvprof} that smaller voids tend to have an averaged velocity profile closer to 0, but we end up with the same conclusion. It is possible that there might exist other approach to prune away spurious voids, but at the same time the pruning method must be supported by theory of voids, or otherwise, the approach is unnatural. An additional pruning criteria could correct the distribution on small scales closer to the SvdW prediction, but it is difficult not to influence voids on other scales at the same time. We conclude that the identified small voids in simulations are unlikely to be spurious, and are indeed slowly collapsing underdense structure.  


\subsection{Environmental Dependence of Void Distribution}
\begin{figure*}
	\includegraphics[scale=0.155]{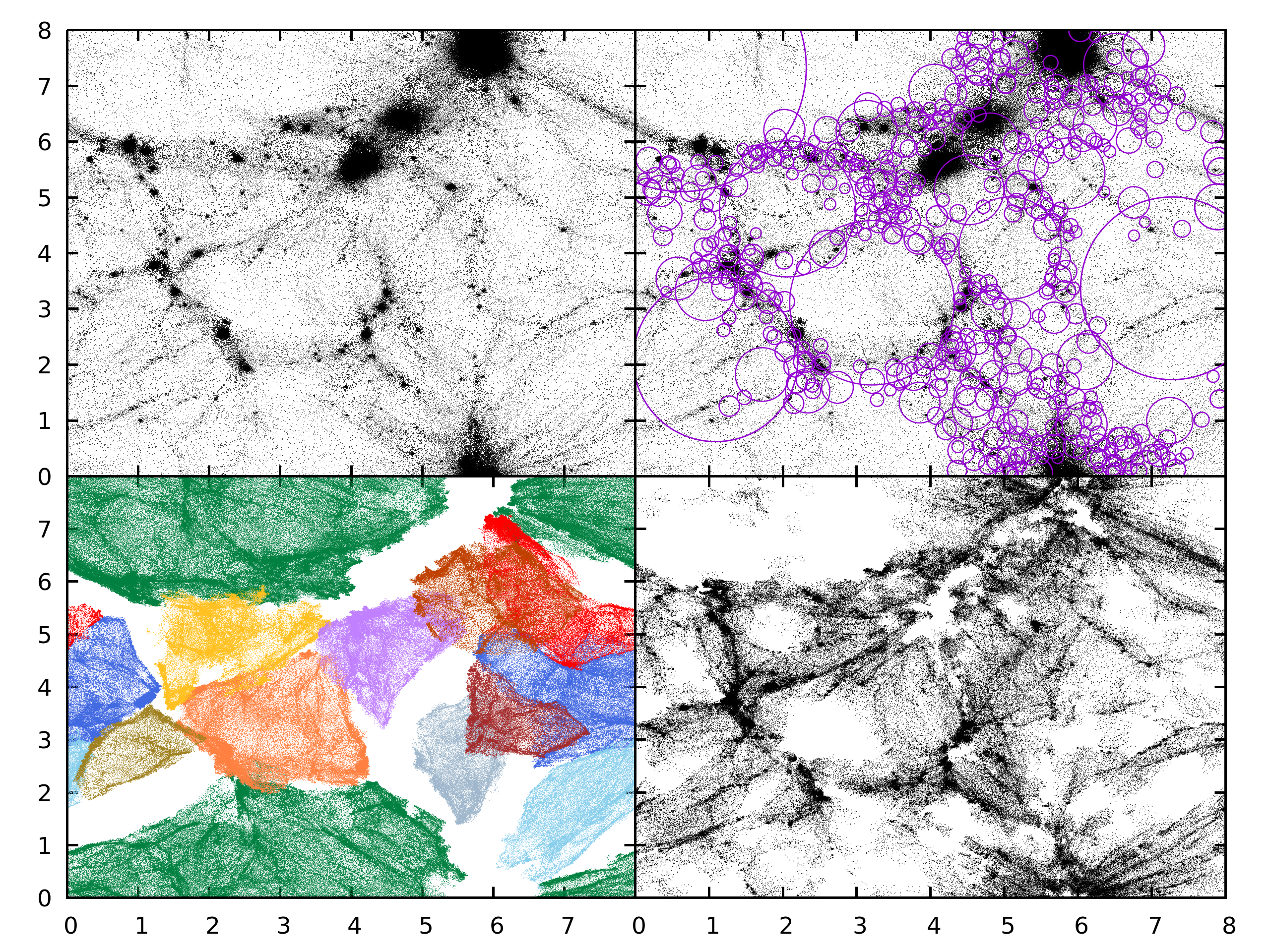}
	\caption{
		Top--left: Dark matter particles in a 8 $\times$  8 $\times$ $[1, 3]\,h^{-1}$ Mpc slice of Phi-0. Top--right:
		Identified spherical voids are denoted as purple circles and superimposed on the dark matter
		distribution. Bottom--left: Member particles of the top eleven largest aspherical voids are shown
		in different colour. Green void is the largest with $R_\text{eff} = 3.8$ $h^{-1}$ Mpc. Bottom--right: Member
		particles of all other small voids are denoted as black dots.
	}
	\label{fig: dmv}
\end{figure*}

Figure \ref{fig: dmv} demonstrates how the dark matter particles, spherical and aspherical voids distribute in a $8 \times 8 \times [1,3]$ slice of Phi--0 simulation. First, it is visually convincing that the spherical large voids locate at the correct underdense regions, and, interestingly, small spherical voids tend to reside close to the filament and overdense regions. The assumption of sphericity can lead to inappropriate inclusion of chunks of filaments passing through the voids, which is unphysical. In the case of aspherical voids, however, the problem disappears. The bottom row of Figure \ref{fig: dmv} shows the distribution of top 11 largest aspherical voids on the left, and all other smaller voids on the right. The border of the aspherical voids are now perfectly aligned with the filament. It is apparent that these large voids share most of the volume of the simulation, but fail to annex overdense regions such as filaments and clusters. However, all the small voids together map out the filamentary structure again suggesting that small voids tend to reside in the filament and avoid the center of clusters, which is in agreement with the result based on spherical voids. Figure \ref{fig: dmv}, therefore, provides us a clear evidence of the existence of small voids, but, more importantly, they tend to reside in overdense structures. 

If it is true that small voids prefer to reside in a preferred environment, we are dealing with the environmental dependence of void formation. Beside visual evidence, we could also provide a quantitative analysis on the environmental dependence by computing the eigenvalues of the tidal tensor, allowing us to determine whether a void is residing in one of these four kinds of tidal environment: cluster, filament, sheet, or a larger void. The method was first suggested by \citet{Hahn2007}, but we follow the alternative procedure in \citet{Paranjape2018}. We first perform Fourier transform on the CIC interpolated density contrast with $128^3$ grid cells, and then inverse Fourier transform the Hessian of the gravitational potential to obtain the tidal tensor. The signs of the eigenvalues of the tidal tensor allow us to classify the void environment. This method contains one free parameter, the smoothing radius, where we will use $R = 0.6 \,h^{-1}$Mpc. The choice of the smoothing radius is to allow measurement of a continuous void size function without unphysical truncations. 
\begin{figure*}
	\includegraphics[scale=0.8]{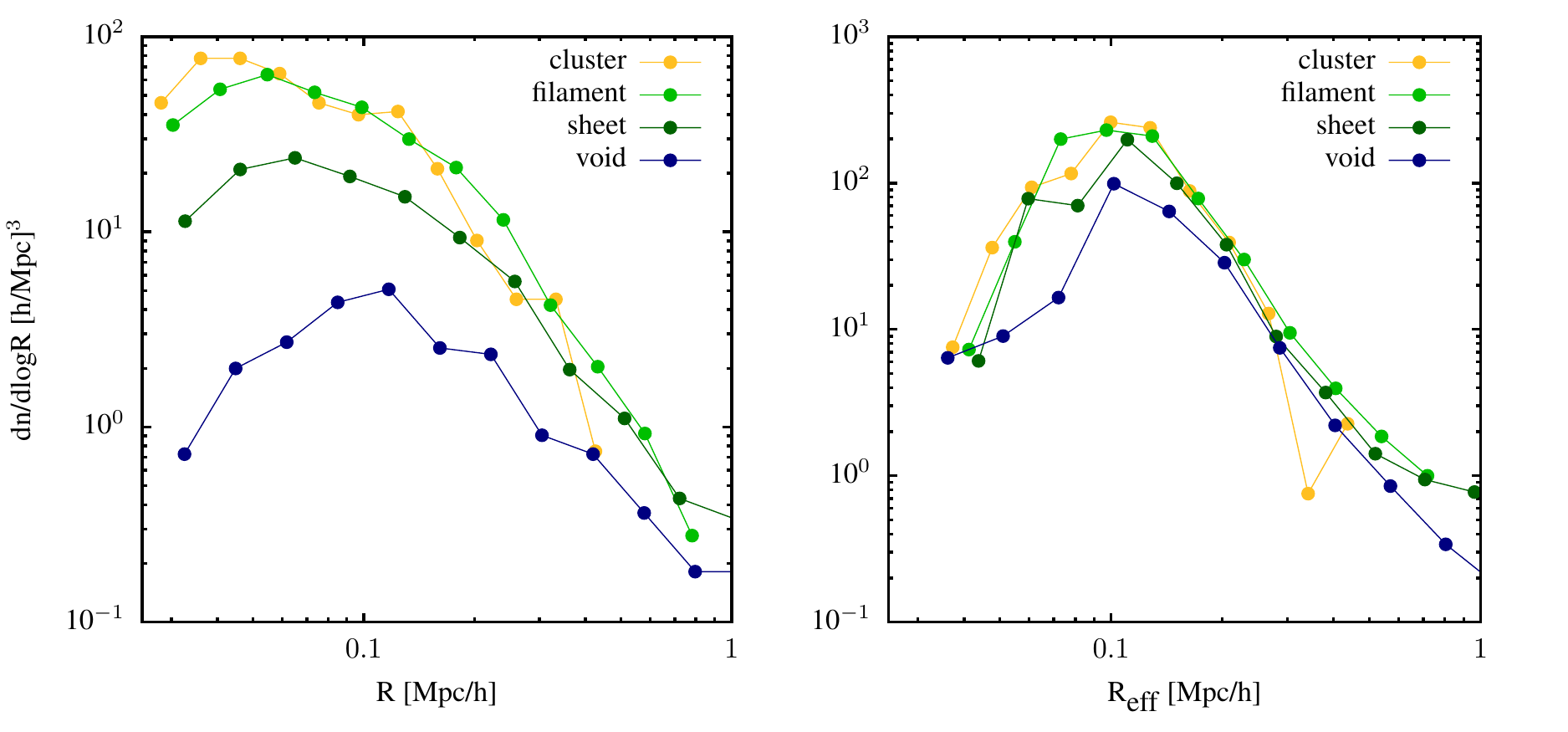}
	\caption{
		Spherical (left) and aspherical (right) void size distributions in different tidal environments in Phi--0 simulation.
	}
	\label{fig: vsfenv}
\end{figure*}
Figure \ref{fig: vsfenv} shows the void size distribution, normalized by their corresponding volume and splited into four different tidal environments for both spherical and aspheircal voids in the Phi--0 simulation. As it shows, a significant amount of voids is indeed residing in filaments, and even in clusters. Also, there is a strong correlation with environment for the case of spherical voids. We propose that the resulting distributions can be explained by the strength of the void--in--void effect. Large underdense regions are occupied by the least amount of voids due to the strong void--in--void effect. The strength weakens in other denser tidal environments, so filaments and clusters experience the weakest void--in--void effect allowing a great number of voids to reside in them. However, if the void--in--cloud process occurs, we should expect the strong void--in--cloud effect in filaments and clusters on the scales below $1 \,h^{-1}$Mpc, leading to strong suppression of formation of small voids, but as Figure \ref{fig: vsfenv} shows, the void density in filaments and clusters is still high relative to that in voids and sheets. All of these suggests that, even in filaments and clusters, the void--in--cloud effect is rather weak at suppressing void distribution when comparing with the void--in--void effect.

The aspherical void distributions show a weaker correlation with environments, but we should emphasize again that the aspherical void distribution is strongly dependent on the zone merging parameter. By decreasing the merging threshold, we can obtain aspherical void distributions with environmental correlation as strong as that of spherical voids. Note that the tails of distribution in cluster decreases faster than all others at $\sim 0.2 \,h^{-1}$Mpc, since the size of cluster becomes too small to contain voids of larger size.

\section{Discussion}
The abundance of voids at current epoch is the result of void formation process, which, as \citet{Sheth2004} suggested, must involve the void--in--void and void--in--cloud effects. Our work has three major conclusions. (i) The measured void distributions reveal an overabundance on small scales $R < 1\,h^{-1}$Mpc, suggesting that the void--in--cloud process might be a rare occurrence. (ii) The velocity profile suggests that most of these small voids are indeed experiencing the void--in--cloud effect, but unlike the SvdW model which predicts complete collapse, they are only partially collapsing underdense structures. (iii) The tidal environment of small voids indicates that the void--in--cloud effect is rather weak when comparing with the void--in--void effect, resulting in the overabundance of voids in filaments and clusters.

We would like to emphasize that this work is the first to provide an in--depth analysis on the void--in--cloud process. Our conclusion provides a concrete evidence for the Eulerian void--in--cloud formulation model \citep{Paranjape2012}, which also proposes the partial collapsing scenario during void formation. With the consideration of correlated steps, they pointed out that it is reasonable to expect the void--in--cloud effect to become sub--dominant; therefore, in qualitative agreement with the overabundance of small voids in simulation. For other modified models, the volume conserving model \citep{Jenning2013} reduces more voids on all scales, so they are not the solutions to the overabundance problem. In \citet{Clampitt2013}, although they took account of the void--in--cloud effect in their modified gravity model, they mentioned that the void--in--cloud effect is uncommon in the real Universe, where our result agrees with their statement. 

The main distinction between halo and void distributions is that the former depends only on one parameter $\delta_\text{c}$ and the latter depends on two parameters $\delta_\text{c}$ and $\delta_\text{v}$. If there is no void--in--cloud effect, the void distribution turns out to be only dependent on one parameter $\delta_\text{v}$; therefore, having a similar behaviour to the halo distribution in the WDM Universe. Since our results suggest a weak void--in--cloud effect, the above case might be true which weakens the uniqueness of using void distribution to probe dark matter types. However, since the void--in--cloud process is not fully non--existent, it is worth studying void statistics in the WDM cosmology in future works.

\section{Conclusion}
We have measured the basic properties of spherical and aspherical voids in $\Lambda$CDM cosmology with Phi-0, Phi-1 and three Multidark-Planck dark matter only simulations. Spherical voids distribution is in agreement with the SvdW model on scales larger than $1\,h^{-1}$Mpc, but inconsistent on smaller scales, where the PS model with the replacement of $\delta_\text{c}$ with $|\delta_\text{v}|$ provides an excellent agreement. For aspherical voids, the number of small voids can be described fully by a naive power law. Both size distributions provide evidences that the void--in--cloud effect is rather weak in contrast to the prediction of the SvdW model in the $\Lambda$CDM Universe. The measured average velocity profile and tidal analysis suggest that small voids are partially collapsing structure, which prefer to reside in overdense filaments and clusters. As a result, the standard definition of the void--in--cloud process, defined by the SvdW model, is an over--simplified version of the process. We could incorporate the partially crushing process in the analytical void model, which has been done by \citet{Paranjape2012}, so it is essential in the future to perform a comparative study with the above modified model. 

It is ambiguous whether a modification to the void finder can resolve the inconsistency. Since our void finder has satisfied the definition of voids based on the excursion set formalism and spherical expansion model, an additional criterion to it becomes artificial and unsupported by any existing theory of voids. We conclude that a modification to the void finder is an unlikely solution.

Several improvements can be considered in future work. Void distribution in different redshifts should also be studied, in order to fully understand the dynamical evolution of small voids and the void--in--cloud process. Although we are only considering dark matter voids in this work, with the consideration of galaxy bias \citep{Furlanetto2005}, it is possible to extend the current studies to galaxy voids which are directly observable in galaxy surveys. 	
\section*{Acknowledgements}
We thank the anonymous referee for constructive suggestions which signficantly improved the manuscript. This work is supported in part by JSPS Grant-in-Aid for Scientific Research (No.17H01101)
and MEXT Grant-in-Aid for Scientific Research (No.17H04828, 18H04334, 18H04337 and 18H05437).
The CosmoSim database used in this paper is a service by the Leibniz-Institute for Astrophysics Potsdam (AIP). The VSMDPL, SMDPL and MDPL2 simulations have been performed on the Supermuc supercomputer at LRZ using time granted by PRACE.
Numerical computations of Phi-0 and Phi-1 simulations were partially carried out on the K computer at
the RIKEN Advanced Institute for Computational Science (Proposal
numbers hp170231, hp180180), Aterui and Aterui II supercomputer at
Center for Computational Astrophysics, CfCA, of National Astronomical
Observatory of Japan.  TI has been supported by MEXT as ``Priority
Issue on Post-K computer'' (Elucidation of the Fundamental Laws and
Evolution of the Universe) and JICFuS.







\end{document}